\documentclass[aps,prx,twocolumn,preprintnumbers,superscriptaddress,longbibliography,10pt]{revtex4-1}
\usepackage[T1]{fontenc} 
\usepackage[utf8]{inputenc}
\usepackage{multirow}
\usepackage{float}
\usepackage{bm}
\usepackage{booktabs} 
\usepackage{array} 
\usepackage{graphicx}
\usepackage{amsmath}
\usepackage{amssymb}
\usepackage{color}
\usepackage[per-mode=symbol,separate-uncertainty]{siunitx} 
\usepackage[capitalize]{cleveref} 
\usepackage{array,multirow,makecell}
\usepackage[table]{xcolor}
\usepackage{soul}
\usepackage{footmisc}

\setcellgapes{1pt}
\makegapedcells
\newcolumntype{R}[1]{>{\raggedleft\arraybackslash }b{#1}}
\newcolumntype{L}[1]{>{\raggedright\arraybackslash }b{#1}}
\newcolumntype{C}[1]{>{\centering\arraybackslash }b{#1}}

\newcommand\footnoteref[1]{\protected@xdef\@thefnmark{\ref{#1}}\@footnotemark}
\normalsize

\DeclareSIUnit{\torr}{torr}

\begin{document}

\author{Luca Planat}
\author{Ekaterina Al-Tavil}
\author{Javier Puertas Mart\'inez}
\author{R\'emy Dassonneville}
\author{Farshad Foroughi} 
\author{S\'ebastien L\'eger}
\author{Karthik Bharadwaj}
\author{Jovian Delaforce}
\author{Vladimir Milchakov}
\author{C\'ecile Naud}
\author{Olivier Buisson} 
\author{Wiebke Hasch-Guichard} 
\author{Nicolas Roch}
\affiliation{Univ. Grenoble Alpes, CNRS, Grenoble INP, Institut N\'eel, 38000 Grenoble, France}
\email{nicolas.roch@neel.cnrs.fr}

\title{Fabrication and characterization of aluminum SQUID transmission lines}

\begin{abstract} 
We report on the fabrication and characterization of \SI{50}{\ohm}, flux-tunable, low-loss, SQUID-based transmission lines.
The fabrication process relies on the deposition of a thin dielectric layer (few tens of nanometers) via Atomic Layer Deposition (ALD) on top of a SQUID array, the whole structure is the covered by a non-superconducting metallic top ground plane. 
We present experimental results from five different samples. We systematically characterize their microscopic parameters by measuring the propagating phase in these structures. We also investigate losses and discriminate conductor from dielectric losses. This fabrication method offers several advantages. First, the SQUID array fabrication does not rely on a Niobium tri-layer process but on a simpler double angle evaporation technique. Second, ALD provides high quality dielectric leading to low-loss devices. Further, the SQUID array fabrication is based on a standard, all-aluminum process, allowing direct integration with superconducting qubits. Moreover, our devices are \textit{in-situ} flux tunable, allowing mitigation of incertitude inherent to any fabrication process. Finally, the unit cell being a single SQUID (no extra ground capacitance is needed), it is straightforward to modulate the size of the unit cell periodically, allowing band-engineering. This fabrication process can be directly applied to traveling wave parametric amplifiers.

\end{abstract}

\maketitle

\section{Introduction}

Being able to reproduce the rich physics of non-linear fiber optics~\cite{Agrawal:1616167} in the microwave domain would be a major milestone in microwave physics, since the nonlinearities at stake in this frequency range are orders of magnitude larger than in the optical domain. Nonlinear optical fibers have brought fiber amplifiers, a key technology for communication systems but are also very appealing to quantum optics since their dispersion can be tailored to create photonic crystals, allowing the investigation of phenomena such as frequency translation of single photons~\cite{McGuinness:2010ja}. In the microwave domain, electrical signals propagate in transmission lines~\cite{pozar2009microwave}. The quantum nature of these microwave photons can usually be discarded, unless the circuits in which they propagate are cooled down to very low temperatures (below \SI{100}{\milli\kelvin}). Under these conditions, we speak about circuit Quantum Electrodynamics (cQED)~\cite{Schoelkopf:2008cs}. While in the optical domain non-linearity can be enhanced by doping fibers with rare-earth materials, microwave superconducting quantum circuits can be made strongly non-linear and low-loss by combining superconducting materials and Josephson junctions~\cite{Anonymous:EB27L8P5} or by taking advantage of the self-non-linearity of disordered superconductors~\cite{Chin:ws}. This approach has been very successful and lead to the observation of strong light-matter coupling~\cite{wallraff2004strong}, resonance fluorescence with extinction as high as 94\%~\cite{Astafiev:2010cm} or near quantum-limited parametric amplifiers~\cite{CastellanosBeltran:2008cg} since it combines both dissipationless and very nonlinear characteristics. However, all these experiments rely on resonant structures -- the microwave equivalent of optical cavities.

So far, only few experiments used nonlinear transmission lines -- the microwave equivalent of nonlinear optical fibers -- in the quantum regime. One notable exception is the demonstration of Traveling Wave Parametric Amplifiers based on either Josephson junctions~\cite{macklin2015near, white2015traveling} or disordered superconductors~\cite{Eom:2012kq, Bockstiegel:2014bta, Vissers:2016bn,Chaudhuri:2017ij}. The small number of experimental implementations is explained by the fact that fabricating a long, low-loss, impedance matched, nonlinear transmission line is very demanding. In the case of a Josephson junction transmission line (JJ-TL), nonlinearity is strong~\cite{Krupko:2018is}, sparing the need for long structures. The challenge here is to lower the naturally large impedance of Josephson junction arrays~\cite{Planat:2019jh} compared to \SI{50}{\ohm} -- by increasing their capacitive effect to the ground, which asks for very concentrated capacitors -- while maintaining sufficient low losses. Combining all these requirements was demonstrated using a complex niobium trilayer fabrication process~\cite{Tolpygo2015Fab}. Disordered superconductor transmission lines, on the other hand, are mainly limited by their relatively weak nonlinearity. Obtaining sizable nonlinear quantum effects requires meter-long structures, which are strongly prone to fabrication defects~\cite{Bockstiegel:2014bta, Adamyan:2016bn}.

\begin{figure}[h]
\includegraphics[width=\linewidth]{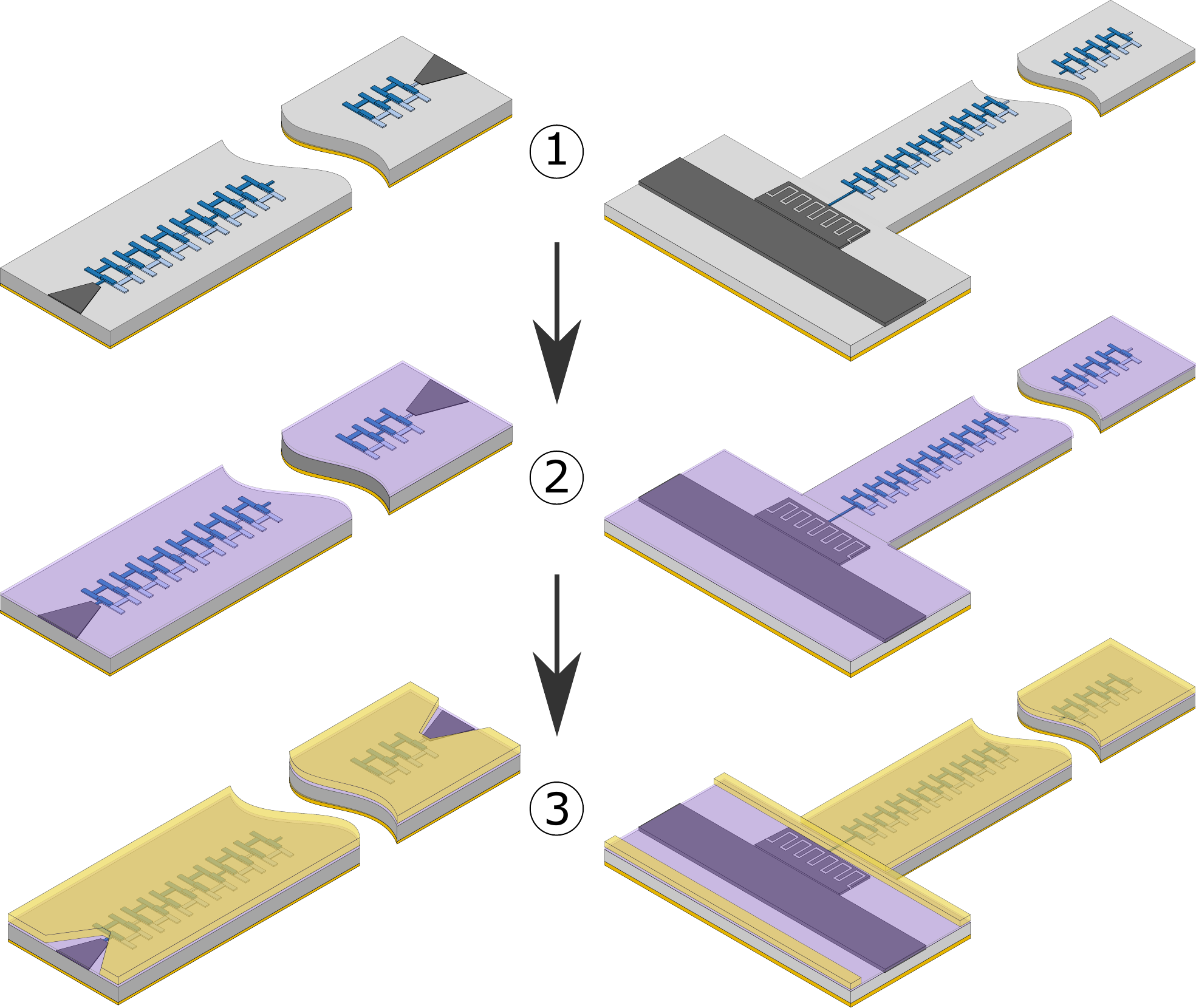}
\caption{\textbf{Fabrication flow for SQUID-based transmission lines and resonators.} Step 1: fabrication of long SQUID arrays using electron-beam lithography and double angle evaporation of aluminium. Step 2: Deposition of a conformal alumina layer via Atomic Layer Deposition (ALD). Step 3: Evaporation of a thick metallic layer (gold or copper) acting as an electrical ground. This layer is patterned using a combination of electron-beam lithography and lift-of. (\textbf{Inset:})}  
\label{fig1}
\end{figure}

In this work, we present a SQUID transmission line (S-TL) fabricated using a simple, aluminium-based process. Impedance matching is obtained via a top ground plane separated from the SQUID array by a very thin alumina layer. These S-TL show losses on-par with previously reported values~\cite{macklin2015near} and a characteristic impedance close to \SI{50}{\ohm}. This impedance can be adjusted in-situ, owing to the flux tunability of the structure~\cite{Jung:2013dr}. Furthermore, thanks to the simple architecture of our S-TL, the impedance of each unit cell can be tailored at will to create photonic-crystal-like transmission lines~\cite{Hutter:2011cj,planat19twpa}. In this article two types of devices, based on the exact same fabrication process, are presented: \SI{50}{\ohm} matched S-TL and resonant structures made of shorter S-TL. The latter are used as test structures to characterise our fabrication process in the single microwave photon regime.

This article is organized as follows. In Section~\ref{fab} we introduce the fabrication flow and the microwave design of the S-TL. Section~\ref{lowtemp} presents the low temperature microwave properties of five different devices. Section~\ref{losses} and Section~\ref{power_dep} focus on the microscopic origins of S-TL losses and their strong power dependance respectively. In the last section magnetic flux response of the S-TL is presented. 

\section{Fabrication process}
\label{fab}

In this section we detail the fabrication process of SQUID-based transmission lines. Four different batches were obtained using this recipe. Five different devices were then characterized at very low temperatures, as reported in \cref{tab1}. Devices are fabricated on high-resistivity silicon substrates (thickness \SI{275}{\micro\meter}). 

\begin{figure}[h]
\includegraphics[width=\linewidth]{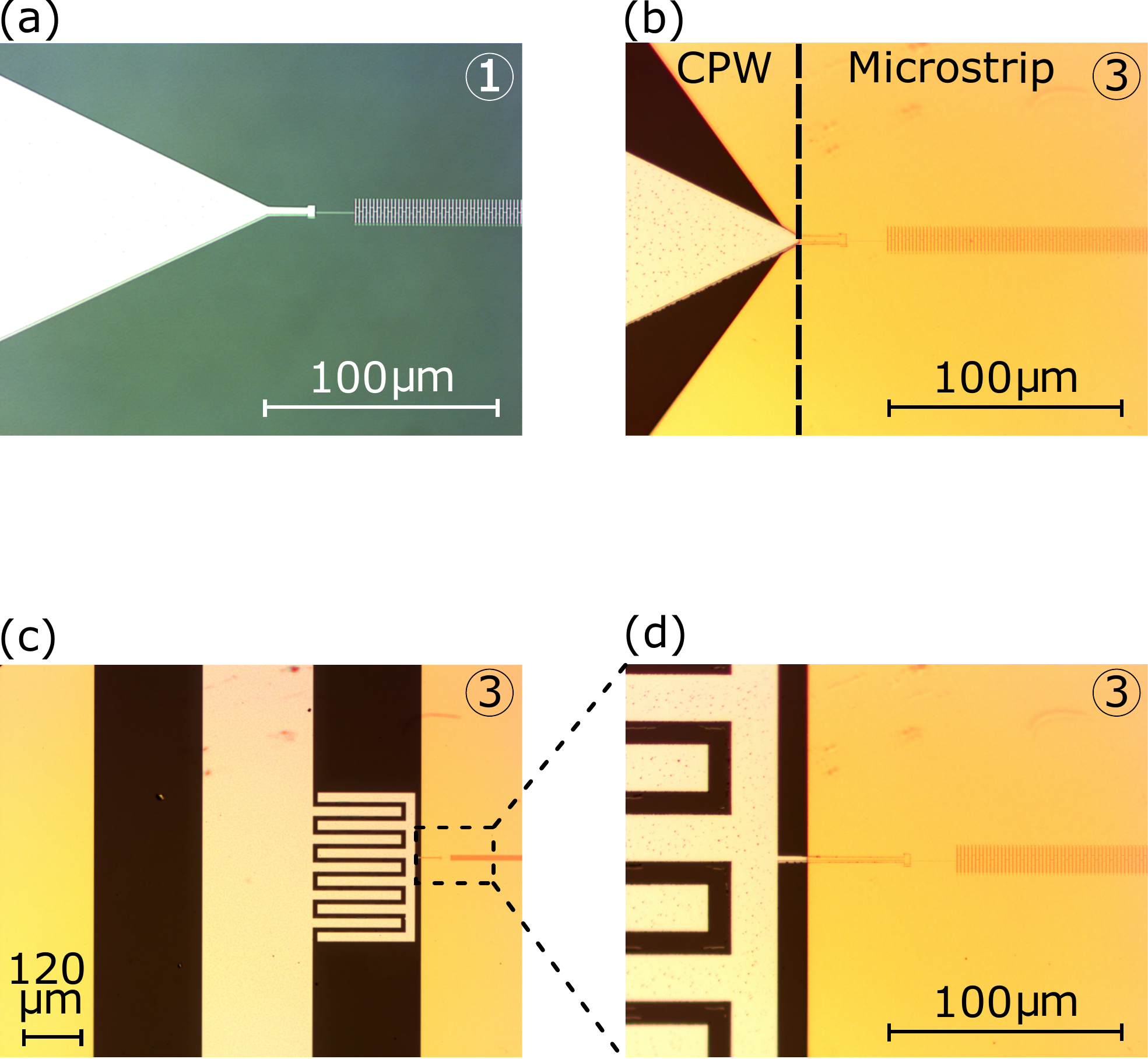}
\caption{\textbf{Pictures of samples from the first batch}. Pictures \textbf{a} and \textbf{b} display SQUID-based transmission lines whereas \textbf{c} and \textbf{d} are showing details of resonant structures used as control samples. Every picture is labeled by a number corresponding to a fabrication step as shown in~\cref{fig1}. (\textbf{a}) Input of the SQUID-based transmission line after the first step. Left side of the picture: bonding pad with tapered shape. Right side of the picture: few dozens of SQUID. (\textbf{b}) Same structure but after step 3. The gold layer is deposited everywhere but on the bonding pad. SQUID are still distinguishable from below the alumina and gold layers. (\textbf{c}) Picture of the resonant structure after step 3. The feed-line is visible in the middle. On the right side: inter-digital capacitor coupling the feed-line to a section of SQUID-based transmission line (623 unit cells). (\textbf{d}) Zoom in.}
\label{fig2}
\end{figure}

The back side of the silicon wafer is covered by a thin layer of titanium (\SI{10}{\nano\meter}, for adhesive purpose) and a thick layer of gold (\SI{200}{\nano\meter}) to ensure good thermal and electrical contact to the sample holder. The fabrication process relies on three simple steps summarised in \cref{fig1}. First a SQUID array is fabricated using double angle evaporation of aluminium (evaporator MEB550S from Plassys), separated by an \textit{in-situ} oxidation to grow the tunnel barrier. For batch 1, 2 and 3, the oxidation pressure is \SI{4}{\torr}, for batch 4 it is \SI{1}{\torr}. The resist mask is patterned with a 100 keV electron beam writer (model nB5 from NanoBeam) to allow bridge free fabrication~\cite{Lecocq:2011dk}. The typical size of a single Josephson junction is ten microns high and half a micron wide (see Appendix~\ref{sec:models}). A unit cell (two Josephson junctions plus connecting wires in a loop) is about \SI{3}{\micro\meter} wide. This technique allows fabrication of low disorder arrays combining up to 2000 unit cells~\cite{PuertasMartinez:2019gk}. However, one of the main difficulty in fabricating such long arrays are stitching errors, coming from a wrong focus of the e-beam due to unavoidable tilt in the substrate. To overcome them, we use the \textit{focus map} feature of our electron beam writer, allowing to readjust dynamically the focus during the writing process. This is achieved by fitting, prior to the writing process, the surface of the chip ($\SI{8.1}{\milli\meter} \times \SI{8.1}{\milli\meter})$ by a tilted plane. The fit is done by measuring the four corners' height where gold marks were previously deposited. The array is terminated by tapered bonding pads (respectively coupling capacitance) as shown in \cref{fig2}\textbf{b} (resp. \textbf{d}). During step 2, a thin film of alumina is deposited via Atomic Layer Deposition (ALD) using the Savannah system, from Cambridge Nanotech. The sample is inserted inside a chamber pumped down to \SI{0.29}{\milli\bar}. We tried two different deposition temperatures, \SI{150}{\celsius} and \SI{200}{\celsius}, to infer the effect of temperature on the dielectric quality. Thicknesses of the various films are reported in \cref{tab1}. ALD films combine low microwave losses and conformal deposition. This latter property is crucial to guarantee electrical isolation between the SQUID and the top ground plane deposited during step 3. Access to the bonding pads is guaranteed by windows in this metallic layer. These openings are obtained via a second lithography step and lift-off of the metallic film. We checked that the thickness of the alumina layer is not affected by this subsequent step. Finally, we could easily contact the bonding pads through the alumina layer using micro-bonding given the thickness of this layer (less than \SI{40}{\nano\meter}).

\begin{table}[ht]
\footnotesize
\begin{ruledtabular}
\begin{tabular}{|C{2.0cm}||C{1.1cm}|C{1.1cm}|C{1.1cm}|C{1.1cm}|C{1.1cm}|}
\hline Device & A & B & C & D & E\\
\hline  Batch & 1 & 2 & 2 & 3 & 4\\
\hline  Type & TL & TL & Res & TL & TL\\
\hline
\hline  ALD temperature (\SI{}{\celsius}) & 150 & 150 & 150 & 200 & 150 \\
\hline  Dielectric thickness (\SI{}{\nano\meter}) & 38 & 38 & 38 & 28 & 28  \\
\hline  Ground thickness (\SI{}{\nano\meter}) & 200 & 400 & 400 & 400  & 1000 \\
\hline  Ground Material & Au & Au & Au & Au  & Cu \\
\hline
\hline  JJ size H(\SI{}{\micro\meter})xW(\SI{}{\micro\meter}) & 12.0 x 0.45 & 10.5 x 0.40 & 10.5 x 0.40 & 12.0 x 0.45 & 12.0 x 0.45 \\
\hline  $L_\text{J}$(\SI{}{\pico\henry})  & 83   & 125 & 125  & 135  & 51  \\
\hline  $C_\text{J}$(\SI{}{\femto\farad}) & 490  & 380  & 380  & 485  & 485  \\ 
\hline  $C_\text{g}$(\SI{}{\femto\farad}) & 34.0 & 26.0 & 31.0 & 46.0 & 50.5 \\
\hline Phase velocity ($10^{6}$ \SI{}{\meter}.\SI{}{\second^{-1}}) & 1.96 & 1.76 & 1.62 & 1.32 & 2.06\\
\hline
\hline  $\tan{\delta}$ ($10^{-3}$)  & 6.5  & 6.0 & 5.0 & 4.0 & 6.5  \\
\hline  $L_\text{c}$ ($10^{-4}$) (\SI{}{\meter^{-1}}.\SI{}{\hertz^{-\frac{1}{2}}}) & 2.5 & 1.0 & x & 0.8 & not measurable  \\
\hline
\hline  Phase calib. Method & Switch & Switch & N/A & Thru & Thru  \\
\hline
\end{tabular}
\end{ruledtabular}
\normalsize
\caption{Summary of the five samples presented in this article. Four are transmission lines (TL) and one is a resonator (Res). Main fabrication characteristics, SQUID characteristics and loss coefficients are summarized. Calibration methods are also reported}  
\label{tab1}
\end{table}

Regarding microwave engineering, we designed the SQUID arrays as a \SI{50}{\ohm} microstrip transmission line. Since our SQUID display a typical inductance $L\approx\SI{100}{\pico\henry}$, and a characteristic impedance $Z_\text{c}=\sqrt{L/C_\text{g}}\approx\SI{50}{\ohm}$ is being sought, it requires a shunt capacitance $C_\text{g}\approx\SI{40}{\femto\farad}$. We model $C_\text{g}$ as a planar capacitance $C_\text{g}=\epsilon_0\epsilon_\text{r}S/t$ (where $\epsilon_0$ is the vacuum permittivity and $S$ the SQUID's area). Given we are using alumina as the insulator (dielectric constant $\epsilon_\text{r}\approx9.8$ ) it translates into a dielectric thickness of $t\approx\SI{30}{\nano\meter}$. These numbers lead to a transmission line with a microstrip geometry with unique features, such as wave velocity below 1\% of the light velocity while being matched to \SI{50}{\ohm} environment.
We also designed a smooth transition from microstrip geometry to Coplanar Waveguide (CPW) geometry to allow for large bonding pads. This transition is shown in~\cref{fig2}.\textbf{b}. We used a conventional tapered shape to keep the impedance constant between the bonding pad and the CPW-microstrip transition. Strictly speaking this is not a CPW geometry since the aluminum layer and the top ground are not exactly in the same plane as they are separated by the alumina layer. Nevertheless, electromagnetic simulations show that the electric field profile is not altered, and that approximating this geometry as a CPW is correct. Indeed alumina and silicon have close dielectric constants and a step of a few tenth of nanometers is negligible compared to the lateral distance between the aluminum and the ground of few hundreds of micron. Regarding resonators, the section of the SQUID-based transmission line (length is approximately 600 unit cells) is capacitively coupled to a feed-line as shown in \cref{fig1} and \cref{fig2}.\textbf{c} and \textbf{d}.

\section{Cryogenic microwave properties}
\label{lowtemp}

\begin{figure}[h]
\includegraphics[width=\linewidth]{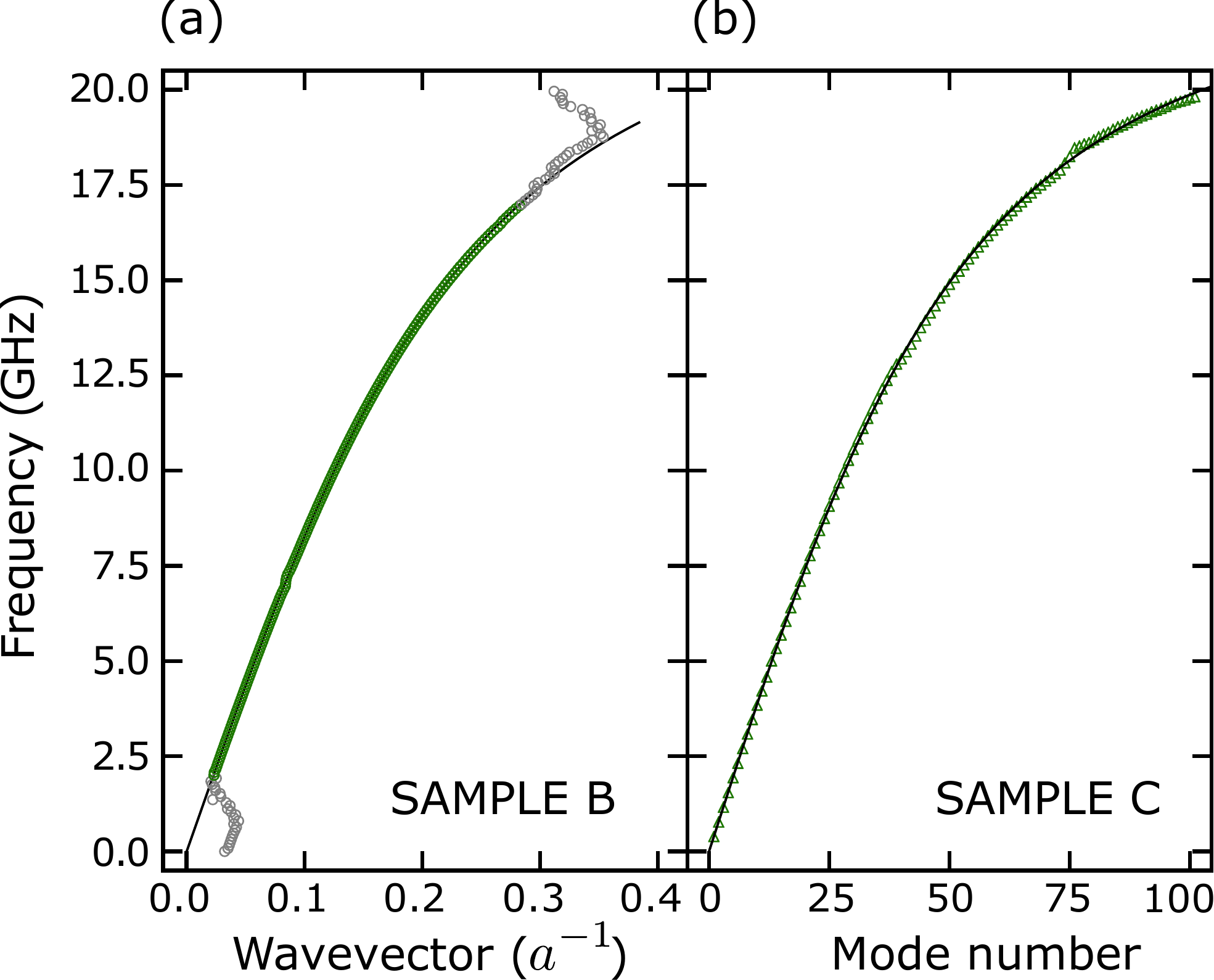}
\caption{\textbf{Dispersion relations} (\textbf{a}) Dispersion relation of a SQUID-based transmission line (sample B). Black line is a fit of the experimental data to \cref{dispersion}. (\textbf{b}) Dispersion relation of a resonant structure (sample C). Data were acquired via two tone spectroscopy and fitted using \cref{dispersion} as well. }
\label{fig3}
\end{figure}

We now turn to the microwave characterization of these samples at very low temperature ($T=\SI{20}{\milli\kelvin} $). Measurements were carried out using a standard cryogenic setup. First we measure the dispersion relation (angular frequency versus wave-vector) of the samples. The experimental protocol to access such quantities is not the same for the transmission lines and the resonant structures. The latter is obtained via two-tone spectroscopy~\cite{Anonymous:2012jo, Weissl:2015do}. The former follows the procedure explained in ~\cite{macklin2015near}. The idea is to measure the propagating phase $\phi$ of a microwave tone along the device under test (DUT). A proper calibration is needed to remove the contribution of the cryogenic measurement setup. To do so, we used two calibration techniques. For samples A and B, a cryogenic microwave switch (model R577433000 from Radiall) would shunt the sample at \SI{20}{\milli\kelvin} to infer, during the same cooldown, both the contribution of the setup and of the DUT. For samples D and E we used a simple thru to calibrate the contribution of the setup. We placed a thru in CPW geometry instead of the actual chip containing the SQUID-based transmission line. This technique requires two different cool-downs but gives a better estimate of the setup contribution (wire bonds are similar to the DUT and there are no extra cables connecting to the switch to the DUT). The last calibration step is to ensure that at zero frequency, this propagating phase is zero. Indeed the phase measured with a Vector Network Analyser (VNA) is defined modulo $2\pi$. We use a procedure similar to Macklin et al. \cite{macklin2015near} to adjust it properly. The propagating phase $\phi$ can then be linked to the wave-vector $k$ using:
\begin{equation}
k(\omega)=\frac{\phi(\omega)}{L},
\label{phase}
\end{equation}
where $L$ is the length of the SQUID array and is perfectly known since both the unit cell size $a$ and the total number of unit cells $N_\text{J}$ are well known. Two measured dispersion relations are plotted in \cref{fig3}. Grey points in \cref{fig3}.\textbf{a} were discarded from the fitting procedure since they are out of the operating frequency range (\SI{3}{\giga\hertz} to \SI{14}{\giga\hertz}) of our cryogenic HEMT amplifier (model LNF-LNC1-12A from Low Noise Factory).

To infer the microscopic parameters of the samples $L_\text{J}$ and $C_\text{g}$ (see Appendix~\ref{sec:models}), we fitted their dispersion relations using the following model~\cite{Weissl:2015do} 
\begin{equation}
\omega_\text{k}=\omega_\text{plasma}(\sqrt{\frac{1-\cos{ka}}{1-\cos{ka}+C_\text{g}/2C_\text{J}}}),
\label{dispersion}
\end{equation}

where the Josephson capacitance $C_\text{J}$ is a fixed parameter and taken as \SI{45}{\femto\farad} per micro meter square~\cite{Fay:2008vi} and $\omega_\text{plasma}$ is the plasma frequency of a SQUID and reads $\omega_\text{plasma}=1/\sqrt{L_\text{J}C_\text{J}}$. 

The values found for $C_\text{g}$ throughout the different batches follow the correct trend: the bigger the junctions are, the bigger $C_\text{g}$ is; the thinner the dielectric is, the bigger $C_\text{g}$ is. Regarding $L_\text{J}$, we experienced a drift from batch 1 to batch 3 towards larger inductance. Junctions from batch 4 were oxidized with a pressure of \SI{1}{\torr} instead of \SI{4}{\torr} in order to compensate this drift. Also, sample B and C should display the same microscopic parameters since they were fabricated within the same batch (see \cref{tab1}). The size of sample B unit cells was modulated periodically while sample C has uniform cells. This modulation is very small (6\% amplitude) and only opens a small photonic gap over a band of few hundreds of megahertz in the dispersion relation. This does not affect its overall shape as shown in~\cref{fig2}, where only a small kink can be seen around \SI{7}{\giga\hertz}. We find the same values for the Josephson inductance but there is a small discrepancy in the ground capacitance which might be due to the fitting procedure or to uniformity in the dielectric layer. Finally, we can define a low frequency ($\omega << \omega_\text{plasma}$) characteristic impedance $Z_\text{c}= \sqrt{L_\text{J}/C_\text{g}}$ for both samples, leading to $Z_\text{c}$ between \SI{65}{\ohm} and \SI{70}{\ohm}.

\section{Conductor and dielectric losses}
\label{losses}

\begin{figure}[h]
\includegraphics[width=\linewidth]{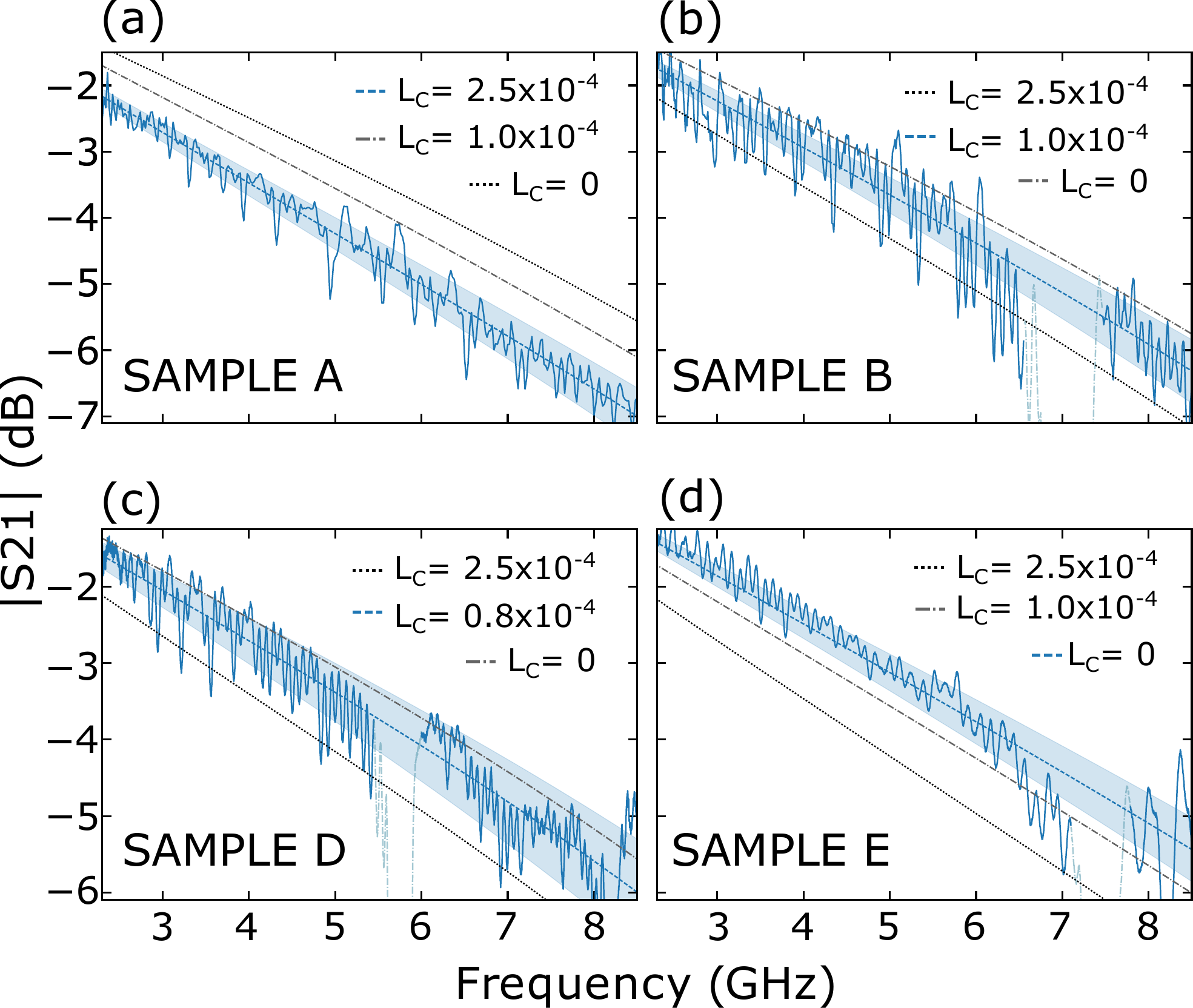}
\caption{\textbf{Calibrated, low temperature transmission of S-TL}. Solid (dashed) lines are experimental data (fits to~\cref{attenuation}). Values of fitting parameters are reported in \cref{tab1}. Shaded blue areas represent $5.10^{-4}$ incertitude on $\tan\delta$. Black dotted and grey dash-dotted lines are obtained using \cref{attenuation} with the same $\tan\delta$ but with different conductor loss (values indicated in each panel in \SI{}{\meter^{-1}\hertz^{-\frac{1}{2}}}). All data were taken with input power $P_\text{in}=-106\ \text{dBm} \pm 3 \text{dBm}$ except for sample A where $P_\text{in}=-101 \text{dBm} \pm 3 \text{dBm}$.}
\label{fig4}
\end{figure}

Losses of the transmission line is another very important parameter. To understand loss mechanisms we probed the calibrated transmission of S-TL, using the calibration technique presented in the previous section. \cref{fig4} shows this calibrated transmissions for samples A, B, D and E. They show losses on the order of \SI{5}{ \decibel/\centi\meter} at \SI{6}{\giga\hertz}. Although these losses are significant, they are comparable to previously reported structure based on Josephson junctions \cite{macklin2015near}. A more thorough study can explain the different loss mechanisms of our devices. Seen as microstrip lines, our samples suffer from two well known phenomena in microwave engineering: dielectric and conductor losses~\cite{gupta1979microstrip}. Within the superconducting community, the latter form of losses have been rightfully neglected since there are no significant losses inside the superconductors nor in thick metallic grounds (superconducting or not). Such assumptions are not valid anymore in our geometry. Although the central conductor is superconducting, the top ground plane is made of a non superconducting metal (gold or copper, for the last version), with a finite thickness (hundreds of nanometers), comparable to the skin depth. This configuration can lead to non negligible conductor losses as we will see later. To determine the origin of losses, we fit the calibrated transmission of the S-TL with a simple model \cite{pozar2009microwave}, where $\alpha_\text{c}$ and $\alpha_\text{d}$ represent the conductor and dielectric losses respectively. The total attenuation of the line can then be expressed as
\begin{equation}
A = e^{(\alpha_{c} + \alpha_\text{d})L},
\label{attenuation}
\end{equation}

Considering our microstrip geometry with very large shunt capacitance, it is safe to assume that the electrical field is mainly confined within the top alumina substrate. Thus, we model the dielectric loss as $\alpha_\text{d}$ as~\cite{pozar2009microwave}:
\begin{equation}
\alpha_\text{d}= \frac{k\tan{\delta}}{2}
\label{alphad}
\end{equation}
We thus have a direct relation between the angular frequency $\omega_\text{k}$ of the signal and the dielectric loss through a free parameter $\tan{\delta}$. In microwave engineering, it is defined as the ratio of the imaginary part over the real part of the complex permittivity of the medium.

Regarding $\alpha_\text{c}$, as mentioned before, we will only take into account top ground loss. 

Given the microstrip line geometry of our system, we consider~\cite{schneider1968microstrip}:
\begin{equation}
\alpha_\text{c}= \frac{R}{Z_\text{0}},
\label{alphac}
\end{equation}
where $Z_\text{0}$ is the characteristic impedance of the line, and $R$ is the \textit{radio resistance} per unit length. $R$ is given by
\begin{equation}
R= \frac{R_\text{s}}{\mu_\text{0}}\frac{\delta l}{\delta n},
\label{radioR}
\end{equation}
where $R_\text{s}=\sqrt{\omega\mu_\text{0}/2\sigma}$ is the skin resistance of the conductor, $\mu_\text{0}$ the vacuum permeability and the ratio $\delta l/\delta n$ characterises the change of inductance per unit length $\delta l$ when the conductor walls recede by a distance $\delta n$. In a metal the current flows within the skin depth $\delta_\text{s}=\sqrt{2/\omega\mu_0\sigma}$. If the conductor is much thicker than $\delta_\text{s}$, the change of inductance is negligible. This amounts to considering the ratio $L\delta l/\mu_0\delta n$ in the order of unity.

On the other hand if the conductor thickness is comparable or smaller than $\delta_\text{s}$, then the change of $\delta l$ is large, the ratio is much larger than one and conductor loss increases. Obtaining a quantitative value of $L\delta l/\mu_\text{0}\delta n$ requires 3D electromagnetic simulations. It is beyond the scope of this paper. However this formula gives the correct order of magnitude for $\alpha_\text{c}$ when the top ground plane is thicker than $\delta_\text{s}$ 

Another important feature is that $\alpha_\text{c}$ should scale as the square root of the frequency. Then we model $\alpha_\text{c}$ as

\begin{equation}
\alpha_\text{c}= L_\text{c}\sqrt{f},
\label{alphac2}
\end{equation}

where f is the signal frequency and $L_\text{c}$ is a prefactor accounting for the conductor losses, which should decrease as the thickness of the top ground metal increases.
Then we fit the attenuation of the various S-TL characterized in this work using two parameters: $\tan\delta$ and $L_\text{c}$. Although S-TL attenuation shows a monotonous trend versus frequency (see \cref{fig4}), we managed to fit independently these two parameters. Indeed surface resistivity depends on the square root of the signal frequency and dielectric loss varies linearly with frequency. In other words, dielectric loss sets the slope of the insertion loss at high frequencies, while conductor loss affects the lower frequencies. All the obtained fitting parameters are presented in \cref{tab1}.

We now turn to the discussion of these parameters. First we measured that $\tan\delta$ remains between $6.5.10^{-3} \pm 5\times10^{-4}$ and $4\times10^{-3} \pm 5\times10^{-4}$ for low-power measurements, which is on par with previously demonstrated JJ-TL~\cite{macklin2015near}. \cref{tab1} summarises the different deposition parameters of alumina. We observe that a higher deposition temperature seems to improve dielectric loss. However, given the incertitude over the loss tangent in this study, we do not claim that deposition temperature is the most important parameters to reduce dielectric loss. However, such values of loss tangent are very promising, since already comparable to state-of-the-art. Moreover, loss tangents as low as $2.45 \times 10^{-3}$, in the single photon regime, was reported for alumina with similar thickness, at very low temperatures~\cite{deng2014characterization}.

Regarding conductor loss, we observe that as the top-ground thickness increases, $L_\text{c}$ drops. We have plotted for what we consider to be the best conductors loss (in dashed blue) for each version as well as other values of conductors (in dashed grey) loss to give insights on the confidence interval (see~\cref{fig4}). As a reminder, batch 2 and 3 should have the same conductor loss since their top-ground thickness is the same. Another interesting point is that a gold conductivity of \SI{100}{\mega\siemens.\meter^{-1}} was reported at very low temperature~\cite{PhysRevLett.77.3885}. This translates into a skin depth $\delta_\text{s}=\SI{600}{\nano\meter}$ at \SI{7}{\giga\hertz}. We can expect the same order for copper ground as gold and copper have close conductivity at room temperature. Batch 1, where the top-ground thickness was well below $\delta_\text{s}$, shows strong conductor loss. In the other batches, with thicker ground planes, losses decreased. Until batch 4 (thickness \SI{1}{\micro\meter}) where conductor loss is below what we can measure~\footnote{If we consider the ratio $L\delta l/\mu_0\delta n$ on the order of unity, the conductor losses $L_\text{c}=\frac{\delta l}{\mu_0\delta n}\frac{\sqrt{\mu_0}}{Z_0\sqrt{2\sigma}}$ reads as $L_\text{c}=\frac{\sqrt{\mu_0}}{LZ_0\sqrt{2\sigma}}$. By taking $Z_0=\SI{50}{\ohm}$, $\sigma=\SI{100}{\mega\siemens\meter^{-1}}$ and $L=\SI{7}{\milli\meter}$, we obtain $L_\text{c}=5.5\times10^{-7}\SI{}{\meter^{-1}\hertz^{-\frac{1}{2} } }$. This justifies our choice to set $L_\text{c}=0$ when fitting sample E.}.
\begin{figure}[h]
\includegraphics[width=\linewidth]{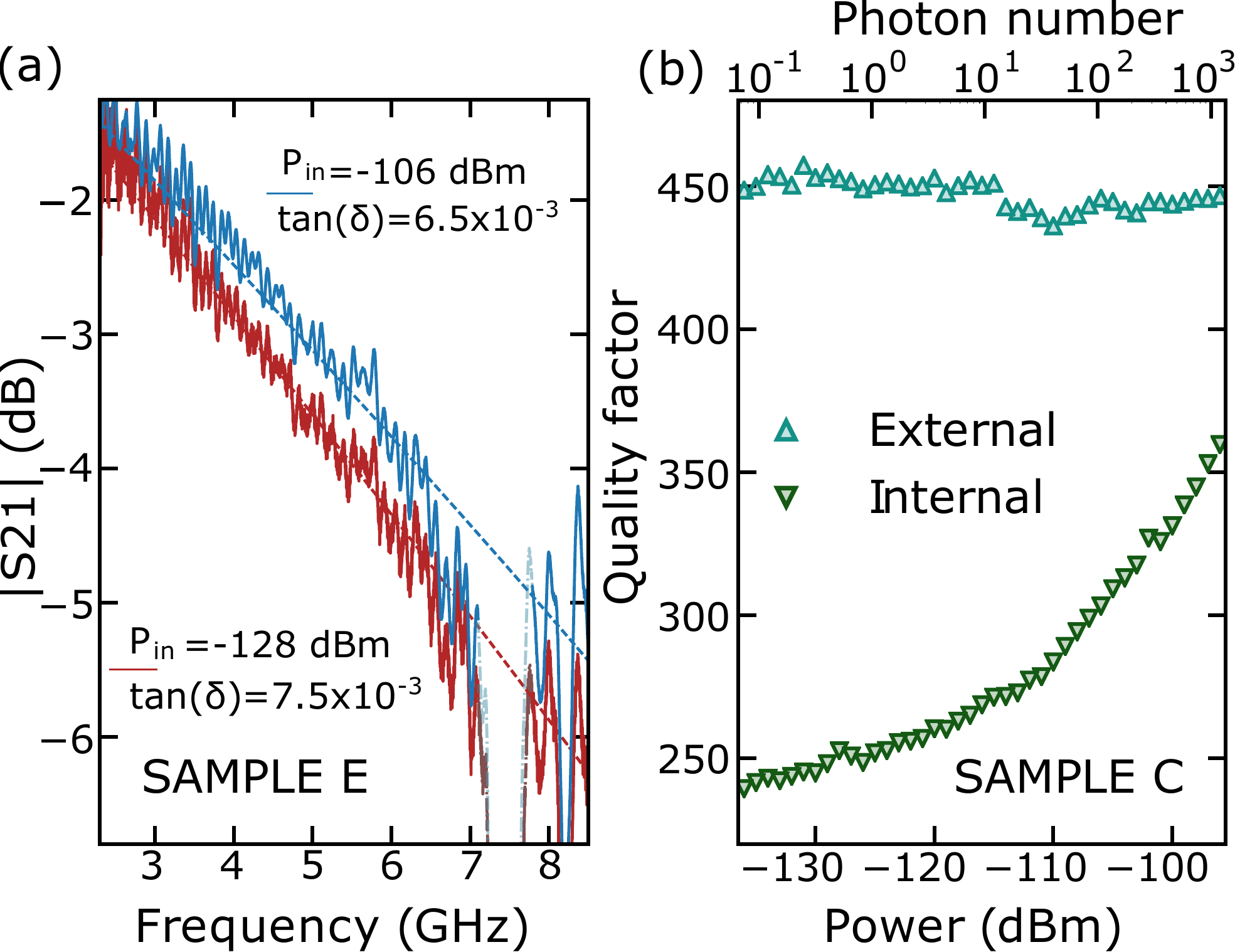}
\caption{\textbf{Power dependence of losses}. For both panels, power is referred to the input of the device. (\textbf{a}) S-TL calibrated transmission (sample E). Input power and fitted loss tangent are reported directly on the figure. (\textbf{b}) Quality factors (internal and external) of a given resonant mode of sample C (angular frequency $\omega_\text{0}=2\pi\times\SI{7.47}{\giga\hertz}$). Fit is obtained for the same resonance at various input powers (see Appendix~\ref{sec:fit}). Photon number (top axis) is obtained using \cref{photon_number}.}
\label{fig5}
\end{figure}

\section{Power-dependent losses}
\label{power_dep}

We now report on the power-dependence of these losses. In both the matched transmission lines and the resonant structures we observe that losses decrease when power increases (\cref{fig5}). The fitted loss tangent of the S-TL presented in \cref{fig5}.\textbf{a} depends on input power and saturates at low power, close to $\tan{\delta}=7.5\times10^{-3} \pm 5\times10^{-4}$. Fitted external and internal quality factors of a given resonant mode are plotted in~\cref{fig5}\textbf{b} for various input powers in the feedline. The external quality factor remains constant while the internal quality factors increases with input power. This observation is in agreement with what was reported in a JJ-TL~\cite{macklin2015near}. At very low powers, close to the single photon regime, losses saturate. This is consistent with the presence of Two-Level-Systems within the dielectric layer~\cite{deng2014characterization}. The internal quality factor and loss tangent are linked via $Q_\text{i} = 1/\tan{\delta}$~\cite{pozar2009microwave}. We can compare quality factors from sample C, which is from the same batch as sample B (calibrated transmission is shown in~\cref{fig4}.\textbf{b}). For sample B, at low signal power, close to the loss tangent saturation, we find $\tan{\delta} = 6.0\times10^{-3} \pm 5\times10^{-4}$. We compare this value to low power quality factors measured for sample C. At $P_\text{input}=-136 \text{dBm}$ \footnote{We estimate the total setup attenuation being \SI{76}{\decibel} $\pm$ \SI{3}{\decibel} as the sum of discrete attenuators (\SI{69}{\decibel}), cable attenuation (\SI{5}{\decibel}) and filter (\SI{2}{\decibel})} ($\bar{n}\approx0.1$) we find, depending on the resonant mode, an internal quality factor between $Q_\text{i,min} = 125 $ and $Q_\text{i,max} = 300 $, with a mean value around 200 (see Appendix~\ref{sec:Qi_f}). This translates into a loss tangent of $\tan{\delta_\text{mean}}=1/200=5.0\times10^{-3}$, in good agreement with the loss tangent found for sample B.

\begin{figure}[htb]
\includegraphics[width=\linewidth]{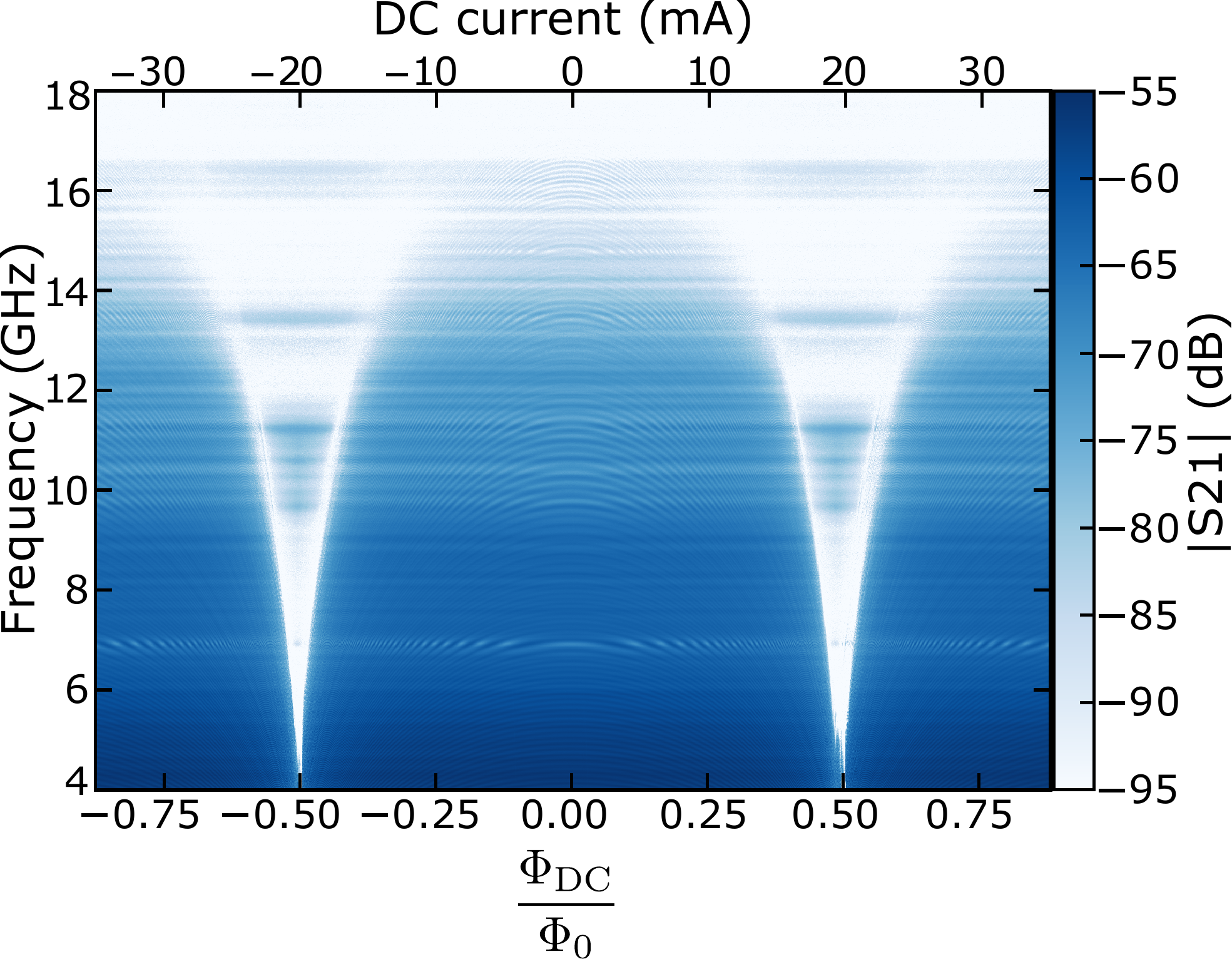}
\caption{\textbf{Transmission of a S-TL versus flux and frequency (sample E).}}
\label{fig6}
\end{figure}

\section{Flux modulation}
\label{flux}

Finally, we describe the flux tunability of a S-TL. \cref{fig6} shows the flux dependence of the transmission of sample E. At low frequencies, below \SI{8}{\giga\hertz}, the transmission is mostly flat with no visible ripples. At higher frequencies one can observe flux-dependent standing waves. This is explained by spurious reflections between the S-TL and its microwave environment. These reflections could be caused by wire-bonds or a non-optimized PCB-to-connector transition. We emphasize the smooth behavior of the device during flux tuning, despite the large number of SQUID. This stability can be attributed to our choice of having a non-superconducting ground plane on top of the SQUID, which prevents flux trapping and effects due to Meissner currents.
Interestingly the plasma frequency of the SQUID can be directly observed as a drop in transmission. At zero flux it is above \SI{20}{\giga\hertz} but drops almost down to zero for magnetic fluxes close to half a fluxoid. 
  
\section{Conclusion}

We have introduced a process to fabricate SQUID transmission lines. It is simple, low-loss and offers in-situ flux tunability. The impedance matching to \SI{50}{\ohm} relies on a simple yet effective idea: a metallic electrical ground deposited on top of the SQUID array, which are separated by a thin alumina layer, guaranteeing electrical isolation. Impedance matching was demonstrated at very low temperatures (\SI{20}{\milli\kelvin}) and very low power (single photon regime). Two loss sources were identified; conductor losses due to the finite thickness of the metallic ground compared to the skin depth and dielectric losses consistent with the presence of two-level systems. We showed that increasing the thickness of the ground plan improves conductor losses. Regarding dielectric losses, we measured loss tangent down to $5.0\times10^{-3}$. This value is comparable to what was reported in the literature~\cite{macklin2015near} but could be improved given the values previously obtained for alumina~\cite{deng2014characterization}. Finally, we demonstrated \textit{in-situ} flux tunability of these S-TL. These devices could be used as Josephson Traveling Wave Parametric Amplifiers, based on four-wave mixing~\cite{macklin2015near,white2015traveling,planat19twpa}. Their flux-tunability offers interesting perspectives regarding three-wave mixing~\cite{Zorin2017TravelingWavePA,Zorin2016JTWPA} or Kerr-free~\cite{Bell2015JTWPA,Zhang2017JTWPA} J-TWPA. These devices could also be used to perform new quantum optics experiments in the microwave domain~\cite{grimsmo2017squeezing}. 

\section*{Acknowledgments}

Very fruitful discussions with D. Basko, K. R. Amin, and A. Ranadive are 
acknowledged.  We also thank M. Selvanayagam for critical reading of the manuscript.  The samples were fabricated in the Nanofab clean room. This research 
was supported by the ANR under contracts CLOUD (project number ANR-16-CE24-0005) and the European Union's Horizon 2020 Research and Innovation Programme, under grant agreement No. 824109, the European Microkelvin Platform (EMP). 
J.P.M. acknowledges support from the Laboratoire d\textquoteright excellence LANEF in Grenoble
(ANR-10-LABX-51-01). R.D. and S.L. acknowledge support from the CFM
foundation and the 'Investisements d'avenir'  (ANR-15-IDEX-02) programs of the French 
National Research Agency.\\

\appendix

\section{Electrical models} 
\label{sec:models}

\begin{figure}[htb]
\includegraphics[width=\linewidth]{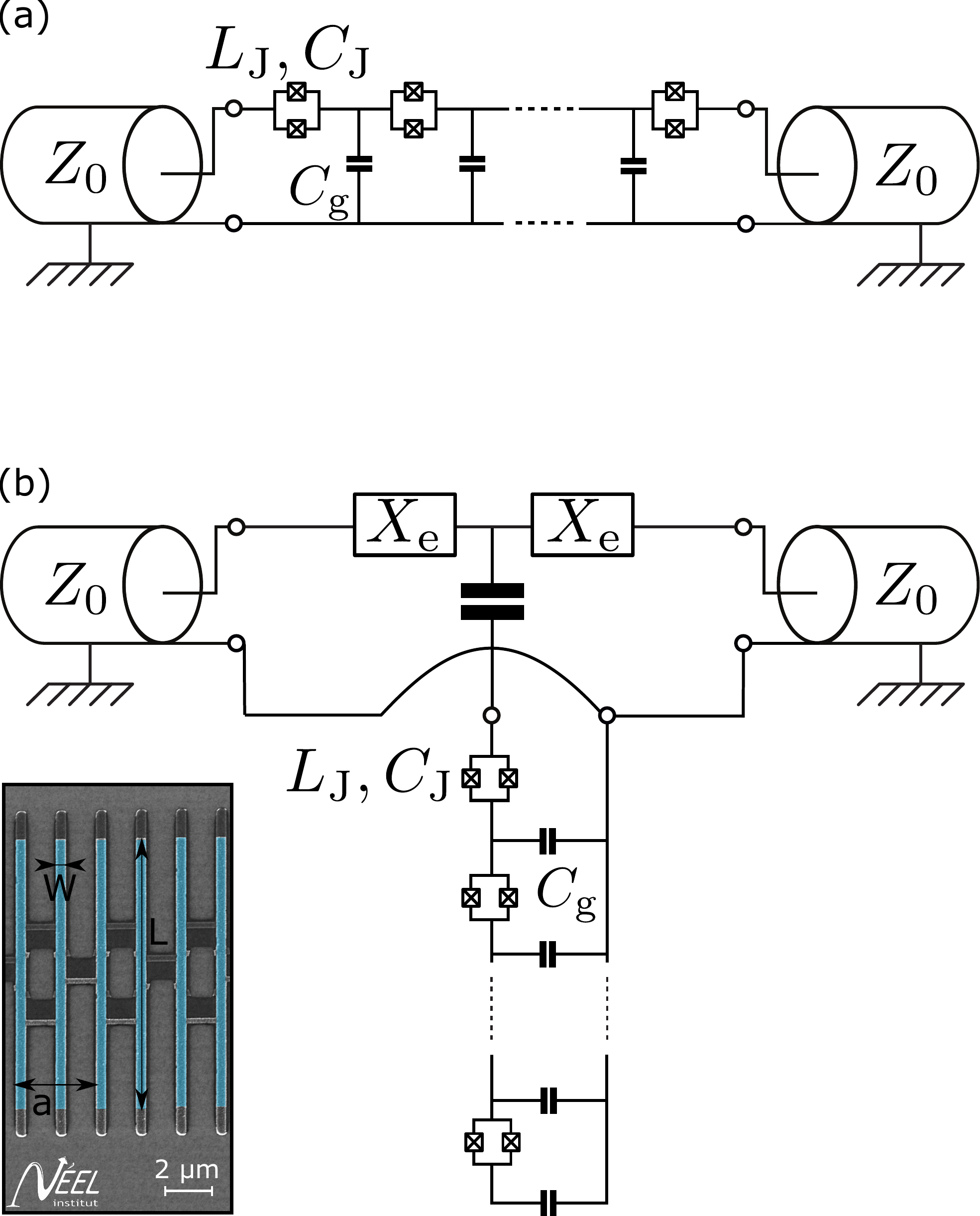}
\caption{Electrical sketches of the two measured structures. (\textbf{a}) Electrical model of the S-TL. (\textbf{b}) Electrical model of the S-Res. \textbf{Inset} SEM picture of 3 SQUID. Highlighted blue regions are Josephson junctions.}
\label{suppmat1}
\end{figure}

In this section, we present the electrical models used to describe the samples we measured. For the S-TL, we used a standard telegrapher model where one SQUID is a cell composed of a nonlinear Josephson inductance $L_\text{J}$, a ground capacitance $C_\text{g}$ and a Josephson capacitance $C_\text{J}$. The electrical sketch is shown in~\cref{suppmat1}.\textbf{a}. The SQUID array is connected to \SI{50}{\ohm} pads via wire bonding.
For the SQUID-based resonator, we use the same model. The only difference is that the line is capacitively coupled to the \SI{50}{\ohm} feedline. An electrical sketch is drawn in \cref{suppmat1}.\textbf{b}. 

\section{S-Res measurements}
\label{sec:fit}

\begin{figure}[htb]
\includegraphics[width=\linewidth]{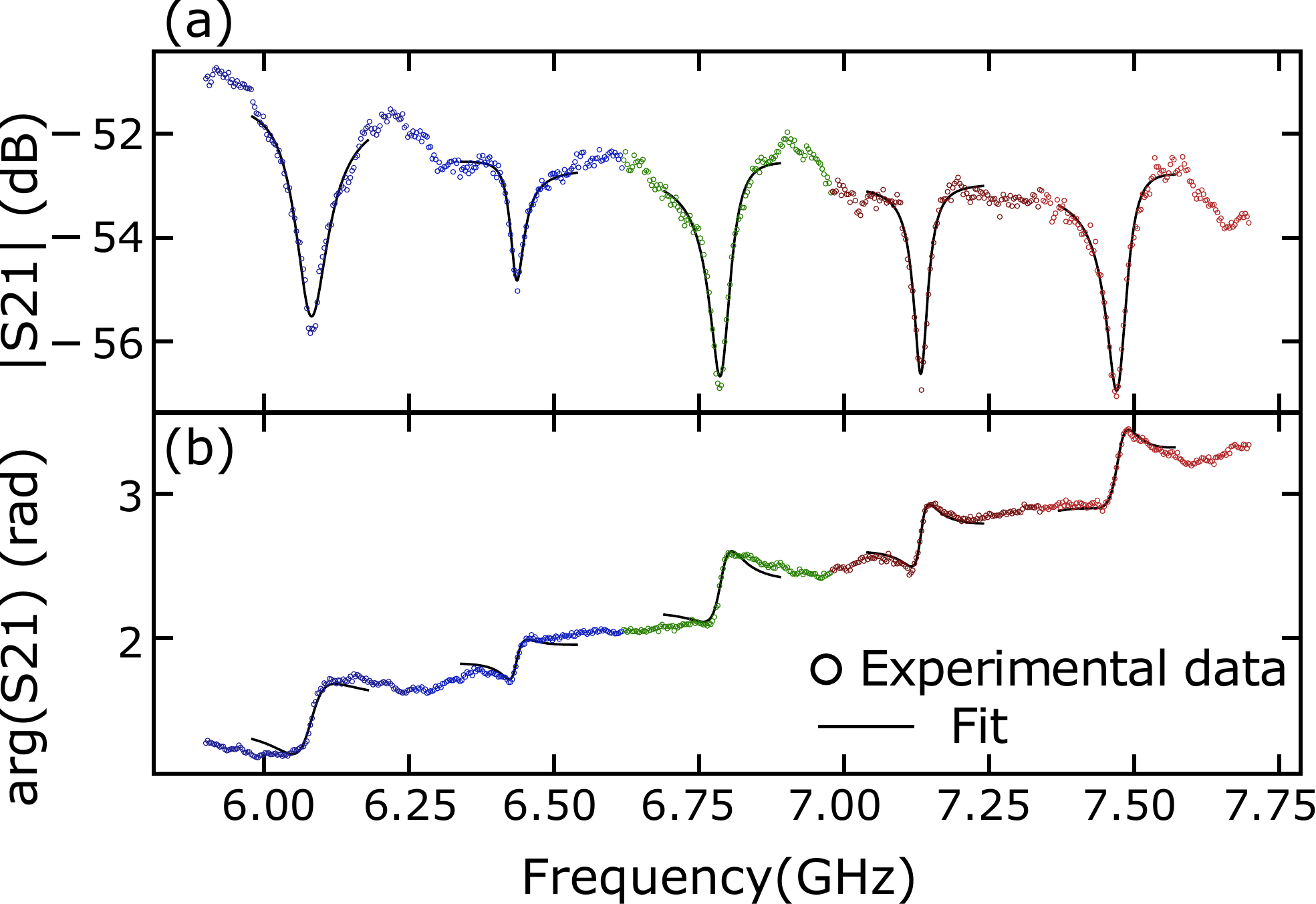}
\caption{Transmission |S21| of the hanger resonators fitted close to the resonances }
\label{suppmat2}
\end{figure}

We present here low temperature measurements of sample C. It is 600 SQUID long with a free spectral range of \SI{350}{\mega\hertz}. Transmission close to resonance is fitted using the formula~\cite{dumur2015}:

\begin{equation} 
\text{S}21 = \frac{Z_\text{0}}{Z_\text{0} + iX_\text{e}} \frac{1+2iQ\text{i}(\frac{\omega - \omega_\text{0}}{\omega_\text{0}})}{1+ \frac{Q_\text{i}}{Q_\text{e}Z_\text{0}(Z_\text{0}+iX_\text{e}}) + 2iQ\text{i}(\frac{\omega - \omega_\text{0}}{\omega_\text{0}})},
\label{resonance}
\end{equation}

where $Z_\text{0}$ is the feedline characteristic impedance, $X_\text{e}$ the reactance from the bonding wires, $\omega_\text{0}$ the resonance frequency of the resonator, $Q_\text{i}$ and $Q_\text{e}$ the internal and external quality factors, respectively.
In~\cref{suppmat2} we present 5 resonances from \SI{6.00}{\giga\hertz} to \SI{7.75}{\giga\hertz} (amplitude and phase) from the S-Res at very low input power. Black lines are fits to \cref{resonance}. From this equation we extract internal and external quality factors (see \cref{fig5}.\textbf{b}) that we plot as a function of input power and photon number inside the cavity. The photon number was calibrated using input/output theory \cite{dumur2015}: 

\begin{equation} 
\bar{n} = \frac{2\frac{\omega_\text{0}}{Q_\text{e}}}{\hbar \omega_\text{0} (\frac{\omega_\text{0}}{Q_\text{e}} + \frac{\omega_\text{0}}{Q_\text{i}})^2} P_\text{input}
\label{photon_number}
\end{equation}

where $\omega_\text{0}$ is the resonant frequency of the cavity and $P_\text{input}$ the signal input power.

\section{Frequency dependency of the internal quality factor}
\label{sec:Qi_f}
\begin{figure}[htb]
\includegraphics[width=\linewidth]{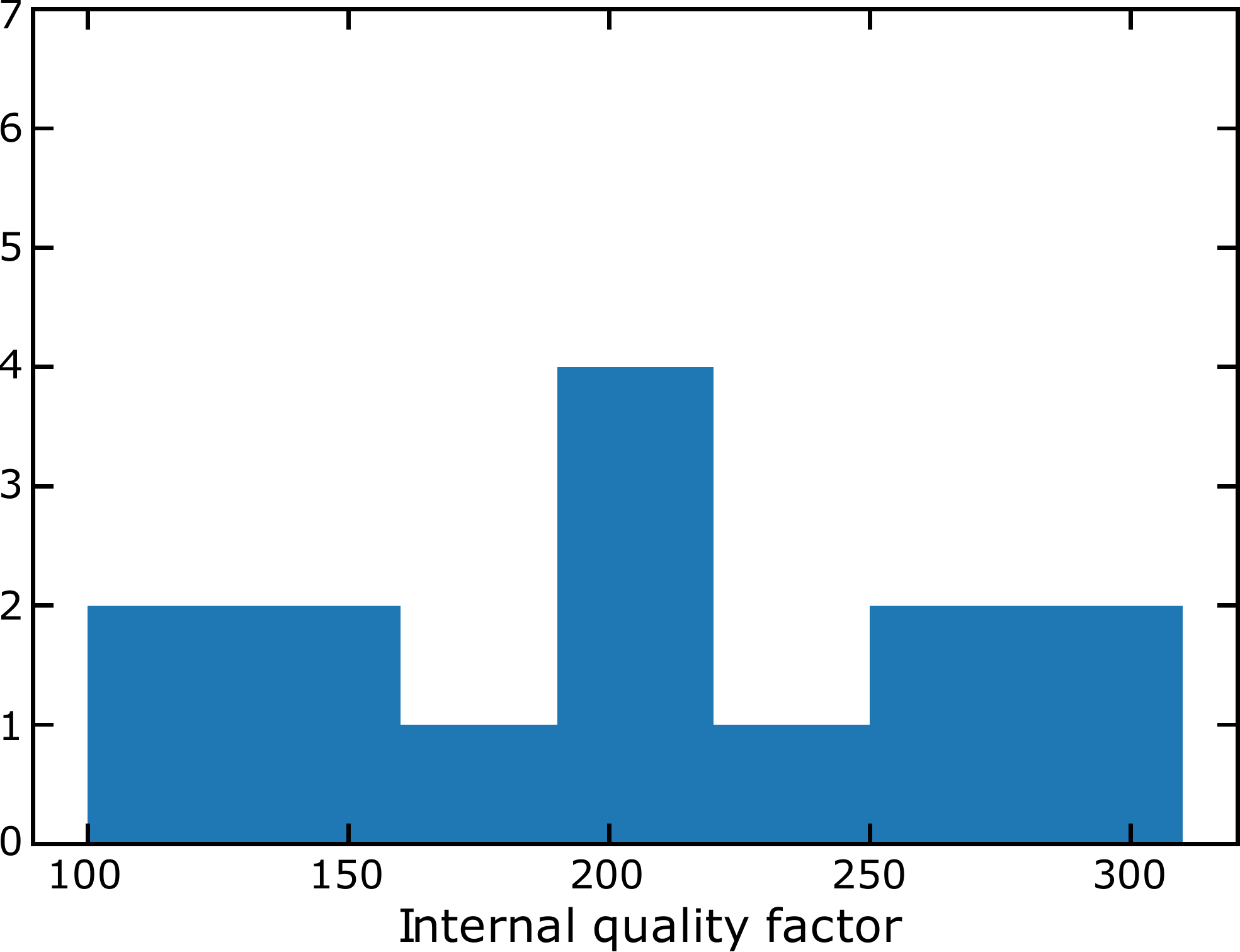}
\caption{Internal quality factors extracted from the fit of 14 modes in sample C between \SI{3.86}{\giga\hertz} and \SI{8.46}{\giga\hertz} at an input power $P_\text{in}=-136\text{dBm} \pm 3\text{dBm}$, corresponding to the single photons level. We observe a spread of the internal quality factor around 200 at this power.}
\label{suppmat3}
\end{figure}

In this section, we present extracted quality factors for very low input power, $P_\text{input} = -136\text{dBm} \pm 3\text{dBm}$, that translates into 0.1 photon (see Appendix \ref{sec:fit} for the photon number calibration). At very low photon number, close to the saturation of the losses, we observe a internal quality factor of 200 with a spread between 125 and 300 for different resonant frequencies as shown in~\cref{suppmat3}.

%

\end{document}